\pdfoutput=1
\documentclass[12pt,a4paper]{article}
\usepackage{ifthen} % for conditional statements
\newboolean{pdflatex}
\setboolean{pdflatex}{true} % False for eps figures 

\newboolean{articletitles}
\setboolean{articletitles}{true} % False removes titles in references

\newboolean{uprightparticles}
\setboolean{uprightparticles}{false} %True for upright particle symbols

\usepackage{longtable} % only for template; not usually to be used in PAPERs
\usepackage{microtype}
\usepackage{lineno}  % for line numbering during review
\usepackage{xspace} % To avoid problems with missing or double spaces after
\usepackage{graphicx}  % to include figures (can also use other packages)
\usepackage{color}
\usepackage{colortbl}
\graphicspath{{./figs/}} % Make Latex search fig subdir for figures

\usepackage{amsmath} % Adds a large collection of math symbols
\usepackage{amssymb}
\usepackage{amsfonts}
\usepackage{upgreek} % Adds in support for greek letters in roman typeset

\newcommand*\patchAmsMathEnvironmentForLineno[1]{%
\expandafter\let\csname old#1\expandafter\endcsname\csname #1\endcsname
\expandafter\let\csname oldend#1\expandafter\endcsname\csname
end#1\endcsname
 \renewenvironment{#1}%
   {\linenomath\csname old#1\endcsname}%
   {\csname oldend#1\endcsname\endlinenomath}%
}
\newcommand*\patchBothAmsMathEnvironmentsForLineno[1]{%
  \patchAmsMathEnvironmentForLineno{#1}%
  \patchAmsMathEnvironmentForLineno{#1*}%
}
\AtBeginDocument{%
\patchBothAmsMathEnvironmentsForLineno{equation}%
\patchBothAmsMathEnvironmentsForLineno{align}%
\patchBothAmsMathEnvironmentsForLineno{flalign}%
\patchBothAmsMathEnvironmentsForLineno{alignat}%
\patchBothAmsMathEnvironmentsForLineno{gather}%
\patchBothAmsMathEnvironmentsForLineno{multline}%
}

\usepackage{hyperref}    % Hyperlinks in references
\usepackage[all]{hypcap} % Internal hyperlinks to floats.

\def\lhcb {LHCb\xspace}
\def\ux85 {UX85\xspace}

\ifthenelse{\boolean{uprightparticles}}%
{

 \def\Ppi         {\ensuremath{\uppi}\xspace}

 \def\Ppsi        {\ensuremath{\uppsi}\xspace}

 \def\PDelta      {\ensuremath{\Delta}\xspace}                 
 \def\PXi      {\ensuremath{\Xi}\xspace}                 
 \def\PLambda      {\ensuremath{\Lambda}\xspace}                 
 \def\PSigma      {\ensuremath{\Sigma}\xspace}                 
 \def\POmega      {\ensuremath{\Omega}\xspace}                 
 \def\PUpsilon      {\ensuremath{\Upsilon}\xspace}                 
 
 \def\PB      {\ensuremath{\mathrm{B}}\xspace}                 
                  
 \def\PD      {\ensuremath{\mathrm{D}}\xspace}

 \def\PJ      {\ensuremath{\mathrm{J}}\xspace}                 
 \def\PK      {\ensuremath{\mathrm{K}}\xspace}

 \def\Pc      {\ensuremath{\mathrm{c}}\xspace}

 \def\Pi      {\ensuremath{\mathrm{i}}\xspace}

}
{

 \def\Ppi         {\ensuremath{\pi}\xspace}

 \def\Ppsi        {\ensuremath{\psi}\xspace}                 
                  
 \mathchardef\PDelta="7101
 \mathchardef\PXi="7104
 \mathchardef\PLambda="7103
 \mathchardef\PSigma="7106
 \mathchardef\POmega="710A
 \mathchardef\PUpsilon="7107
                  
 \def\PB      {\ensuremath{B}\xspace}                 
                  
 \def\PD      {\ensuremath{D}\xspace}

 \def\PJ      {\ensuremath{J}\xspace}                 
 \def\PK      {\ensuremath{K}\xspace}

 \def\Pc      {\ensuremath{c}\xspace}

 \def\Pi      {\ensuremath{i}\xspace}

}

\def\c     {\ensuremath{\Pc}\xspace}

\def\pion  {\ensuremath{\Ppi}\xspace}

\def\pip   {\ensuremath{\pion^+}\xspace}

\def\kaon  {\ensuremath{\PK}\xspace}
  \def\Kbar  {\kern 0.2em\overline{\kern -0.2em \PK}{}\xspace}

\def\Kz    {\ensuremath{\kaon^0}\xspace}
\def\Kzb   {\ensuremath{\Kbar^0}\xspace}
\def\KzKzb {\ensuremath{\Kz \kern -0.16em \Kzb}\xspace}
\def\Kp    {\ensuremath{\kaon^+}\xspace}
\def\Km    {\ensuremath{\kaon^-}\xspace}

\def\KpKm  {\ensuremath{\Kp \kern -0.16em \Km}\xspace}

  \def\Dbar    {\kern 0.2em\overline{\kern -0.2em \PD}{}\xspace}
\def\D       {\ensuremath{\PD}\xspace}

\def\Dz      {\ensuremath{\D^0}\xspace}
\def\Dzb     {\ensuremath{\Dbar^0}\xspace}
\def\DzDzb   {\ensuremath{\Dz {\kern -0.16em \Dzb}}\xspace}
\def\Dp      {\ensuremath{\D^+}\xspace}
\def\Dm      {\ensuremath{\D^-}\xspace}

\def\DpDm    {\ensuremath{\Dp {\kern -0.16em \Dm}}\xspace}

\def\B       {\ensuremath{\PB}\xspace}
  \def\Bbar    {\kern 0.18em\overline{\kern -0.18em \PB}{}\xspace}

\def\Bs      {\ensuremath{\B^0_s}\xspace}

\def\Bc      {\ensuremath{\B_c^+}\xspace}

\def\jpsi     {\ensuremath{{\PJ\mskip -3mu/\mskip -2mu\Ppsi\mskip 2mu}}\xspace}

  \def\Y#1S{\ensuremath{\PUpsilon{(#1S)}}\xspace}% no space before {...}!

\def\L {\ensuremath{\PLambda}\xspace}

\def\BR         {\BF}
\def\to                 {\ensuremath{\rightarrow}\xspace}

\def\AT#1     {\ensuremath{A_T^{#1}}\xspace}           % 2

\def\C#1      {\ensuremath{\mathcal{C}_{#1}}\xspace}                       % 9
\def\Cp#1     {\ensuremath{\mathcal{C}_{#1}^{'}}\xspace}                    % 7
\def\Ceff#1   {\ensuremath{\mathcal{C}_{#1}^{\mathrm{(eff)}}}\xspace}        % 9  
\def\Cpeff#1  {\ensuremath{\mathcal{C}_{#1}^{'\mathrm{(eff)}}}\xspace}       % 7
\def\Ope#1    {\ensuremath{\mathcal{O}_{#1}}\xspace}                       % 2
\def\Opep#1   {\ensuremath{\mathcal{O}_{#1}^{'}}\xspace}                    % 7

\newcommand{\tev}{\ensuremath{\mathrm{\,Te\kern -0.1em V}}\xspace}
\newcommand{\gev}{\ensuremath{\mathrm{\,Ge\kern -0.1em V}}\xspace}
\newcommand{\mev}{\ensuremath{\mathrm{\,Me\kern -0.1em V}}\xspace}
\newcommand{\kev}{\ensuremath{\mathrm{\,ke\kern -0.1em V}}\xspace}
\newcommand{\ev}{\ensuremath{\mathrm{\,e\kern -0.1em V}}\xspace}
\newcommand{\gevc}{\ensuremath{{\mathrm{\,Ge\kern -0.1em V\!/}c}}\xspace}
\newcommand{\mevc}{\ensuremath{{\mathrm{\,Me\kern -0.1em V\!/}c}}\xspace}
\newcommand{\gevcc}{\ensuremath{{\mathrm{\,Ge\kern -0.1em V\!/}c^2}}\xspace}
\newcommand{\gevgevcccc}{\ensuremath{{\mathrm{\,Ge\kern -0.1em V^2\!/}c^4}}\xspace}
\newcommand{\mevcc}{\ensuremath{{\mathrm{\,Me\kern -0.1em V\!/}c^2}}\xspace}

\newcommand{\stat}{\ensuremath{\mathrm{(stat)}}\xspace}
\newcommand{\syst}{\ensuremath{\mathrm{(syst)}}\xspace}

\newcommand{\chisq}{\ensuremath{\chi^2}\xspace}

\def\gsim{{~\raise.15em\hbox{$>$}\kern-.85em
          \lower.35em\hbox{$\sim$}~}\xspace}
\def\lsim{{~\raise.15em\hbox{$<$}\kern-.85em
          \lower.35em\hbox{$\sim$}~}\xspace}

\def\PDF {PDF\xspace}
\def\ptot       {\mbox{$p$}\xspace}

\def\evtgen     {\mbox{\textsc{EvtGen}}\xspace}
\def\pythia     {\mbox{\textsc{Pythia}}\xspace}

\def\geant      {\mbox{\textsc{Geant4}}\xspace}

\def\photos     {\mbox{\textsc{Photos}}\xspace}

\def\tell1  {TELL1\xspace}
\def\ukl1   {UKL1\xspace}

\newcommand{\etal}{{\slshape et al.\/}\xspace}

\usepackage{cite} % Allows for ranges in citations
\usepackage{mciteplus}

\begin{document}

%%%%%%%%%%%%%%%%%%%%%%%%%
%%%%% Title     %%%%%%%%%
%%%%%%%%%%%%%%%%%%%%%%%%%
\renewcommand{\thefootnote}{\fnsymbol{footnote}}
\setcounter{footnote}{1}

% %%%%%%% CHOOSE TITLE PAGE--------
%\onecolumn
% \input{title-LHCb-ANA}
%\input{title-LHCb-CONF}
% $Id: title-LHCb-PAPER.tex 10646 2011-10-12 13:51:38Z uegede $
% ===============================================================================
% Purpose: LHCb-PAPER journal paper title page template
% Author: 
% Created on: 2010-09-25
% ===============================================================================

%%%%%%%%%%%%%%%%%%%%%%%%%
%%%%%  TITLE PAGE  %%%%%%
%%%%%%%%%%%%%%%%%%%%%%%%%
\begin{titlepage}
\pagenumbering{roman}

% Header ---------------------------------------------------
\vspace*{-1.5cm}
\centerline{\large EUROPEAN ORGANIZATION FOR NUCLEAR RESEARCH (CERN)}
\vspace*{1.5cm}
\hspace*{-0.5cm}
\begin{tabular*}{\linewidth}{lc@{\extracolsep{\fill}}r}
\ifthenelse{\boolean{pdflatex}}% Logo format choice
{\vspace*{-2.7cm}\mbox{\!\!\!\includegraphics[width=.14\textwidth]{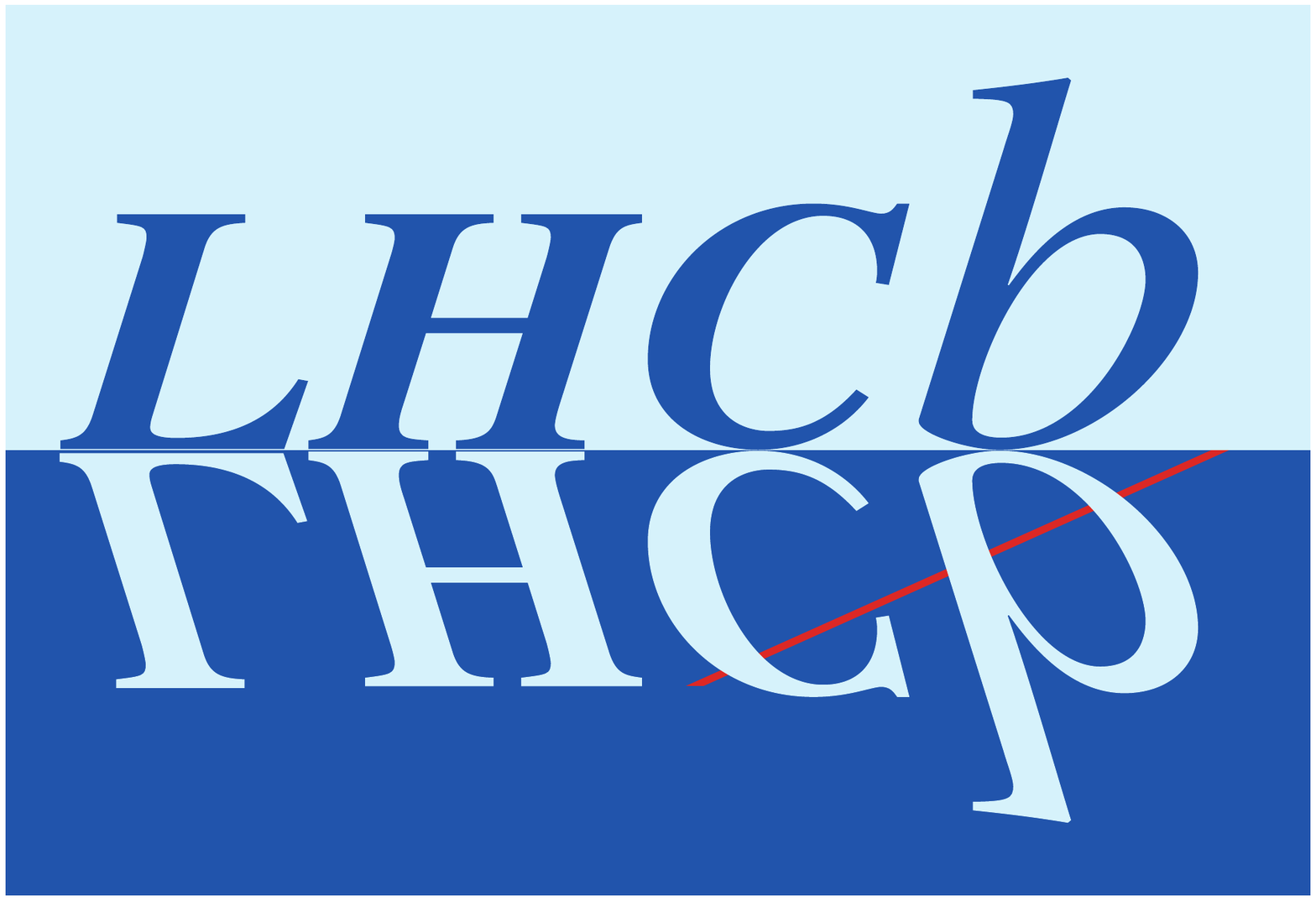}} & &}%
{\vspace*{-1.2cm}\mbox{\!\!\!\includegraphics[width=.12\textwidth]{lhcb-logo.eps}} & &}%
\\
 & & CERN-PH-EP-2014-163 \\  % ID 
 & & LHCb-PAPER-2014-025 \\  % ID 
 & & 8 July 2014 \\ %\today \\ % Date - Can also hardwire e.g.: 23 March 2010
 & & \\
% not in paper \hline
\end{tabular*}

\vspace*{4.0cm}

% Title --------------------------------------------------
{\bf\boldmath\huge
\begin{center}
Measurement of the ratio of 
$B_c^+$ branching fractions \\
to $J/\psi \pi^+$ and $J/\psi \mu^+ \nu_\mu$  
final states
\end{center}
}

\vspace*{1.0cm}

% Authors -------------------------------------------------
\begin{center}
The LHCb collaboration\footnote{Authors are listed at the end of this paper.}
\end{center}

\vspace{\fill}

% Abstract -----------------------------------------------
\begin{abstract}
  \noindent
The first measurement that relates semileptonic and hadronic decay rates of the $B_c^+$ meson is performed 
using proton-proton collision data corresponding to
$\rm 1.0~fb^{-1}$ of integrated luminosity collected with the LHCb detector.
The measured value of the ratio of branching fractions,
${\cal B}(B_c^+ \to J/\psi \pi^+)/{\cal B}(B_c^+\to J/\psi\mu^+\nu_\mu)=0.0469 \pm  0.0028\, \stat \pm 0.0046\, \syst$,
is at the lower end of available theoretical predictions.

\end{abstract}

\vspace*{1.0cm}

\begin{center}
  Submitted to Physical Review D
\end{center}

\vspace{\fill}

{\footnotesize 
\centerline{\copyright~CERN on behalf of the \lhcb collaboration, license \href{http://creativecommons.org/licenses/by/4.0/
}{CC-BY-4.0}.}}
\vspace*{2mm}

\end{titlepage}

%%%%%%%%%%%%%%%%%%%%%%%%%%%%%%%%
%%%%%  EOD OF TITLE PAGE  %%%%%%
%%%%%%%%%%%%%%%%%%%%%%%%%%%%%%%%

%  empty page follows the title page ----
\newpage
\setcounter{page}{2}
\mbox{~}
%\newpage

% Author List ----------------------------
%  You need to get a new author list!
%\input{LHCb_authorlist.tex}

\cleardoublepage

%\twocolumn
% %%%%%%%%%%%%% ---------

\renewcommand{\thefootnote}{\arabic{footnote}}
\setcounter{footnote}{0}

%%%%%%%%%%%%%%%%%%%%%%%%%%%%%%%%
%%%%%  Table of Content   %%%%%%
%%%%%%%%%%%%%%%%%%%%%%%%%%%%%%%%
%%%% Uncomment next 2 lines if desired
%\tableofcontents
%\cleardoublepage

%%%%%%%%%%%%%%%%%%%%%%%%%
%%%%% Main text %%%%%%%%%
%%%%%%%%%%%%%%%%%%%%%%%%%

\pagestyle{plain} % restore page numbers for the main text
\setcounter{page}{1}
\pagenumbering{arabic}

%% Uncomment during review phase. 
%% Comment before a final submission.
%\linenumbers

% You can include short sections directly in the main tex file.
% However, for larger papers it is desirable to split the text into
% several semiautonomous files, which can be revised independently.
% This is especially useful when developing a document in
% collaboration with several people, since then different parts can be
% edited independently.  This type of file organization is shown here.
% 

\newboolean{prl}
\setboolean{prl}{false} % False for eps figures 

\newboolean{figsdir}
\setboolean{figsdir}{false} % use figs/ for figures

\newlength{\figsize}
\setlength{\figsize}{0.8\hsize}
% --------------------
\def\bcjpopp{B_c^+\to\jpsi\pi^+[\pi^-\pi^+]}
\def\jpopp{\jpsi\pi^+[\pi^-\pi^+]}
\def\bcjppp{B_c^+\to\jpsi\pi^+\pi^-\pi^+}
\def\bc2sp{B_c^+\to\psi(2S)\pi^+}
\def\bu2sp{B^+\to\psi(2S)\pi^+}
\def\p2sppj{\psi(2S)\to\pi^+\pi^-\jpsi}
\def\bujp{B^+\to\jpsi\pi^+}
\def\budppp{B^+\to \bar{D}^{*0} \pi^+\pi^-\pi^+}
\def\budp{B^+\to \bar{D}^{*0} \pi^+}
\def\bujppp{B^+\to\jpsi\pi^+\pi^-\pi^+}
\def\bujkpp{B^+\to\jpsi K^+\pi^-\pi^+}
\def\bujk{B^+\to\jpsi K^+}
\def\cospj{\cos(\pi,\jpsi)}
\def\r31{R_{3/1}}
\def\BR{{\cal B}}
\def\L{{\cal L}}
\def\DLL{\Delta_{\rm sig/bkg}({\rm -2ln\L})}
\def\PDF{{\cal P}}
\def\NDOF{\hbox{\rm ndf}}
\def\R31{\BR(\bcjppp)/\BR(\bcjp)}
% -------------------------
\def\R{{\cal R}}
\def\jp{\jpsi\pi}
\def\jm{\jpsi\mu}
\def\mjm{m_{\jpsi\mu}}
\def\mjp{m_{\jpsi\pi}}
\def\tmjm{\bar{m}_{\jpsi\mu}}
\def\tmjp{\bar{m}_{\jpsi\pi}}
\def\bcjp{B_c^+\to\jpsi\pi^+}
\def\bcjk{B_c^+\to\jpsi K^+}
\def\bcjm{B_c^+\to\jpsi\mu^+\nu_\mu}
\def\Bctojm{\bcjm}
\def\Yfeed{f}
\def\psifeed{\psi_{\it f}}
\def\Rf{R_{\it f}}
\def\RfJ{R_{{\it f\,J}}}
\def\Bcasc{\BR_{\rm casc\,{\it f}}}
\def\BcascJ{\BR_{\rm casc\,{\it f\,J}}}
\def\Reps{R_{\epsilon\,{\it f}}}

\noindent

\section{Introduction}

The $B_c^+$ meson is the ground state of the $\bar{b}c$ quark-pair 
system and is the only meson in which weak-interaction decays of both constituents 
compete with each other.\footnote{Charge-conjugate states are implied in this article.} 
About 70\% of the decay width
is expected to be due to the $c \to s$ transition,
favored by the Cabibbo-Kobayashi-Maskawa quark-coupling hierarchy~\cite{Gouz:2002kk}. 
This decay process has recently been observed in the $\Bc\to \Bs\pip$ mode \cite{LHCb-PAPER-2013-044}.
The complementary $b \to c$ transition, which is predicted to account for 20\%\ of the decay width,
is more straightforward to observe experimentally, 
having a substantial probability to produce a \jpsi meson.
Among such decays, semileptonic $\Bc\to\jpsi\ell^+\nu_{\ell} ~(\ell = \mu, e)$ 
and hadronic $B_c^+ \to \jpsi \pi^+$ channels have played
a special role in many measurements.
The semileptonic decays were used in the discovery of the $B_c^+$ meson \cite{Abe:1998wi},
the measurements of its
lifetime \cite{Abe:1998wi,Abulencia:2006zu,Abazov:2008rba,LHCb-PAPER-2013-063}
and the measurement of the production cross-section at the Tevatron \cite{Abe:1998wi}.
The $B_c^+ \to \jpsi \pi^+$ decays were used to measure its lifetime \cite{Aaltonen:2012yb}, 
mass \cite{Aaltonen:2007gv,Abazov:2008kv,LHCB-PAPER-2012-028},
production cross-section at the LHC \cite{LHCB-PAPER-2012-028}
and as a reference for other hadronic branching fraction 
measurements \cite{LHCb-PAPER-2011-044,LHCb-PAPER-2012-054,LHCb-PAPER-2013-010,LHCb-PAPER-2013-021,LHCb-PAPER-2013-047,LHCb-PAPER-2014-009}.
However, there is no experimental determination of the relative size of semileptonic and hadronic decay rates.
The goal of this work is a measurement of the ratio of branching fractions,
\begin{equation}
\R\equiv\frac{\BR(B_c^+ \to \jpsi \pi^+)}{\BR(\Bc\to\jpsi\mu^+\nu_{\mu})},
\end{equation}
and to test various theoretical models of $B_c^+$ meson decays,
for which predictions of $\R$ 
vary over a wide range, 0.050--0.091 \cite{Ref3CC,Ref4ANSS1,*Ref4AKNT2,Ref7EMV,Ref5CF,Kiselev,Ebert,Ref1IKS,RefKLL}.

\section{Analysis outline}

Final states containing a muon offer a distinctive
experimental signature and can be triggered and reconstructed
with high efficiency at LHCb. Therefore, this analysis relies on $\jpsi$ decays to $\mu^+\mu^-$.
Since the neutrino is not detected, both of the studied decay modes 
are reconstructed using a $\jpsi$ candidate plus a charged track ($t^+$), referred to as the {\it bachelor} track.
The mass of $\jpsi\pi^+$ signal candidates peaks at the $B_c^+$ mass within the experimental resolution, 
allowing a straightforward signal yield extraction
in the presence of relatively small backgrounds under the signal peak.
The main challenge in this analysis is the signal yield extraction for the $\Bc\to\jpsi\mu^+\nu_{\mu}$
decay mode, as the $\jpsi\mu^+$ mass ($\mjm$) distribution is broad due to the undetected neutrino. 
To suppress the dominant backgrounds, the analysis is restricted 
to the $\mjm>5.3 \gev$ endpoint 
region and uses the mass-shape difference 
between the signal and the remaining background to extract the $\Bc\to\jpsi\mu^+\nu_{\mu}$ signal 
yield.\footnote{Units in which $c=1$ are used.}
In this mass region the neutrino has low energy, thus 
the $B_c^+\to\jpsi\mu^+\nu_\mu$  candidates are kinematically 
similar to the $B_c^+\to\jpsi\pi^+$ candidates. 
Therefore, many reconstruction uncertainties cancel in the ratio of their rates,
allowing a precise measurement of $\R(\mjm>5.3 \gev)$.
This endpoint value is then extrapolated to the full phase space 
using theoretical predictions.
Since the $B_c^+$ and $\jpsi$ are both $1S$ heavy quarkonia states, 
the form factors involved in predicting the extrapolation factor and 
the shape of the mass distribution at the endpoint
have only modest model dependence.

\section{Detector and data sample}

The analysis is performed on a data sample of $pp$ collisions at a
center-of-mass energy of 7~TeV,  collected during 2011 by the LHCb experiment  
and corresponding to an integrated luminosity of 1.0\,fb$^{-1}$. The \lhcb detector~\cite{Alves:2008zz} is a single-arm forward
spectrometer covering the \mbox{pseudorapidity} range $2<\eta <5$,
designed for the study of particles containing $b$ or $c$
quarks. The detector includes a high-precision tracking system
consisting of a silicon-strip vertex detector surrounding the $pp$
interaction region~\cite{LHCb-DP-2014-001}, a large-area silicon-strip detector located
upstream of a dipole magnet with a bending power of about
$4{\rm\,Tm}$, and three stations of silicon-strip detectors and straw
drift tubes~\cite{LHCb-DP-2013-003} placed downstream of the magnet.
The tracking system provides a measurement of momentum, \ptot,  with
a relative uncertainty that varies from 0.4\% at low momentum to 0.6\% at 100$\gev$.
The minimum distance of a track to a primary vertex, the impact parameter (IP), 
is measured with a resolution of $(15+29/p_{\rm T})\mu$m,
where $p_{\rm T}$ is the component of \ptot transverse to the beam, in $\gev$.
Different types of charged hadrons are distinguished using information
from two ring-imaging Cherenkov detectors~\cite{LHCb-DP-2012-003}. Photon, electron and
hadron candidates are identified by a calorimeter system consisting of
scintillating-pad and preshower detectors, an electromagnetic
calorimeter and a hadronic calorimeter. Muons are identified by a
system composed of alternating layers of iron and multiwire
proportional chambers~\cite{LHCb-DP-2012-002}.

Simulated event samples are generated for the signal decays and 
the decay modes contributing to the background.
In the simulation, $pp$ collisions are generated using
\pythia~\cite{Sjostrand:2006za}  with a specific \lhcb
configuration~\cite{LHCb-PROC-2010-056}.  The production of 
\Bc mesons, which is not adequately simulated in \pythia, is performed 
by the dedicated generator \mbox{\textsc{Bcvegpy}}\cite{Chang:2003cq}. 
Several dynamical models are
used to simulate $\Bctojm$ decays.
Decays of hadronic particles
are described by \evtgen~\cite{Lange:2001uf}, in which final-state
radiation is generated using \photos~\cite{Golonka:2005pn}. 
The interaction of the generated particles with the detector and its
response are implemented using the \geant
toolkit~\cite{Allison:2006ve, *Agostinelli:2002hh} as described in
Ref.~\cite{LHCb-PROC-2011-006}.

\section{Data selection}

This analysis relies on 
$\jpsi t^+$ candidates satisfying the trigger~\cite{LHCb-DP-2012-004}, 
which consists of a
hardware stage, based on information from the muon
system, followed by a two-level software stage, which applies a full event
reconstruction. At the hardware stage, a muon with $p_{\rm T}>1.5 \gev$, 
or a pair of muons with $\sqrt{p_{\rm T\,1}p_{\rm T\,2}}>1.3 \gev$, is required.  
The subsequent lower-level software triggers 
require a charged-particle track with $p_{\rm T}>1.7 \gev$ 
($p_{\rm T}>1.0 \gev$ if identified as muon) 
and with an IP relative to any primary $pp$-interaction vertex (PV) larger than $100$~$\mu$m.
A dimuon trigger, which requires a large dimuon mass, $m_{\mu^+\mu^-}>2.7 \gev$, and each muon to have $p_{\rm T}>0.5 \gev$,
 complements the single track triggers.
The final software trigger stage requires  either  a $\jpsi\to\mu^+\mu^-$ candidate 
with a $\jpsi$ decay vertex separation from the nearest PV of at least 
three standard deviations, or that a two- or three-track combination, which includes a muon, 
is identified as a secondary vertex using a multivariate selection \cite{LHCb-DP-2012-004}.

In the offline analysis, 
$\jpsi\to\mu^+\mu^-$ candidates are selected with the following criteria: 
$p_{\rm T}(\mu)>0.9 \gev$;
$p_{\rm T}(\jpsi)>1.5 \gev$;   
$\chi^2$ per degree of freedom ($\NDOF$) for the two muons to form a common vertex $\chi^2_{\rm vtx}(\mu^+\mu^-)/\NDOF<9$;
and a mass consistent with the $\jpsi$ meson. 
The separation of the $\jpsi$ decay vertex from the nearest PV must be at least 
five standard deviations.    
The bachelor track, and at least one of the muons from the decay of the $\jpsi$ meson,
 must not point to any PV, through the requirement $\chi^2_{\rm IP}>9$. 
The quantity $\chi^2_{\rm IP}$ is defined as the
difference between the \chisq of the PV fitted with and
without the considered particle. 
The bachelor track must not be collinear within 0.8$^\circ$
with either of the muons from the $\jpsi$ meson decay 
and must satisfy $p_{\rm T}>0.5 \gev$ ($>1.0 \gev$ for $\pi^+$). 
A loose kaon veto is applied to the pion candidates, $\ln[\L(K)/\L(\pi)]<5$, 
where $\L$ is the particle identification likelihood \cite{arXiv:1211-6759}.
The  $\jpsi$ candidates are combined  with the bachelor tracks in a kinematic fit to form 
$B_c^+$ candidates with the known $\jpsi$ mass and the $B_c^+$ vertex used as constraints.   
The $B_c^+$ candidate must satisfy $\chi^2_{\rm vtx}(\jpsi t^+)/\NDOF<9$ %, $p_{\rm T}(B_c^+)>4$ \gev 
and have a pseudo-proper decay time 
greater than $0.25$~ps. 
The pseudo-proper decay time is determined as $L\cdot m_{\jpsi t}/|\vec{p}_{\jpsi t}|$, 
where $L$ is the projection of the distance between the $\Bc$ production and decay vertices 
onto the direction of the $\jpsi t^+$ momentum $\vec{p}_{\jpsi t}$
and $m_{\jpsi t}$ is the $\jpsi t^+$ mass.

Four discriminating variables ($x_i$) are
used in a likelihood ratio to improve the background suppression. 
Three of the variables are common
between the two channels:
$\chi^2_{\rm vtx}(\jpsi t^+)/\NDOF$,
$\chi^2_{\rm IP}(B_c^+)$,
and the cosine of the angle between the $\jpsi$ meson and 
the bachelor track transverse momenta.
The latter quantity peaks at positive values for the signal as the $B_c^+$ meson has a high transverse momentum.
Background events in which particles are combined 
from two different $B$ decays usually peak at negative values,
whilst those due to random combinations of particles are more uniformly distributed.  
The $\chi^2_{\rm IP}(B_c^+)$ variable is small for $B_c^+\to\jpsi\pi^+$ decays since the $B_c^+$ momentum 
points back to the PV. 
For $B_c^+\to\jpsi\mu^+\nu_{\mu}$ candidates, the pointing is only approximate 
since the neutrino is not reconstructed. 
However, $\chi^2_{\rm IP}(B_c^+)$ is often smaller than for the background events 
because the neutrino has low momentum.
The fourth variable for the $\jpsi\pi^+$ mode is $\chi^2_{\rm IP}(t^+)$, while for
the $\jpsi\mu^+\nu_\mu$ mode it is the pseudo-proper decay time,
as $\chi^2_{\rm IP}(t^+)$ is found to be ineffective for this channel.
The four one-dimensional signal probability density functions (PDFs), $\PDF_{\rm sig}(x_i)$, 
are obtained from a simulated 
sample of signal events.
The background PDFs, $\PDF_{\rm bkg}(x_i)$, are obtained from the data 
in the $B_c^+\to\jpsi\pi^+$ mass sidebands (5.35--5.80 and 6.80--8.50 $\gev$)
and from the simulation of inclusive backgrounds from $B_{u,d,s}\to\jpsi X$ decays
($X$ denotes one or more particles) for the $B_c^+\to\jpsi\mu^+\nu_{\mu}$ candidates.  
The requirement $\DLL = -2 \sum_{i=1}^4 \ln[\PDF_{\rm sig}(x_i)/\PDF_{\rm bkg}(x_i)] < 1.0$ $(<0.0)$ 
preserves about 93\% (87\%) of signal events for 
$B_c^+\to\jpsi\pi^+$ ($B_c^+\to\jpsi\mu^+\nu_{\mu}$ with $\mjm>5.3 \gev$) 
and efficiently suppresses the backgrounds.
These requirements minimize the expected average statistical uncertainty on the signal yields,
given the observed background levels in each channel. 

\section{Extraction of the $\bcjp$ signal} 
\label{sec:bcjp}

An extended maximum likelihood fit
to the unbinned distribution of observed $\mjp$ values  
yields $N_{\jp}=839\pm40$ $\bcjp$ signal events and 
is shown in Fig.~\ref{fig:mjp}.
\ifthenelse{\boolean{prl}}{
\begin{figure}[tbhp]
}{
\begin{figure}[t]
}
  \begin{center}
  \ifthenelse{\boolean{pdflatex}}{
    \ifthenelse{\boolean{figsdir}}{
       \includegraphics*[width=\figsize]{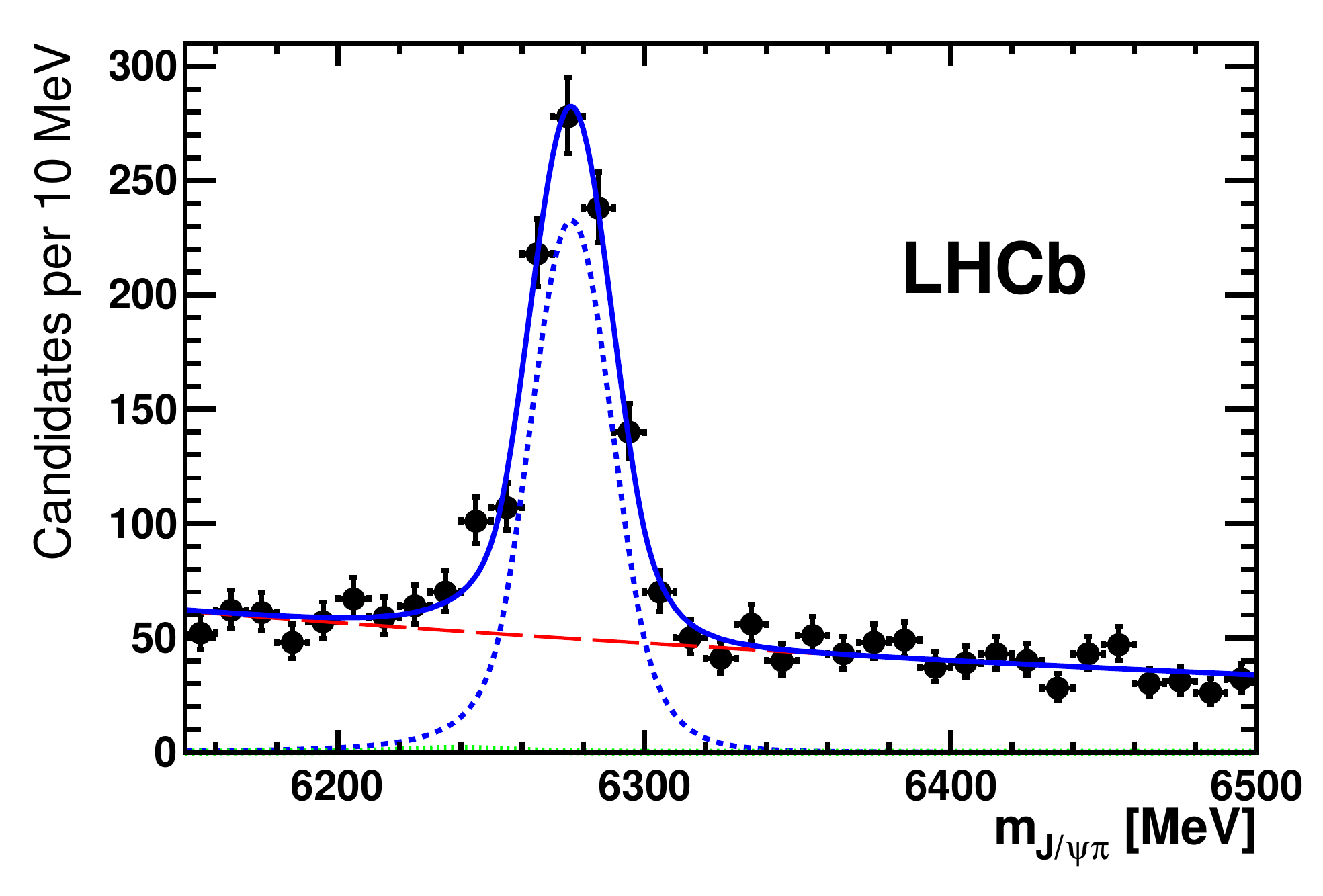}
    }{
       \includegraphics*[width=\figsize]{JpsiPiMassFit.pdf}
    }
   }{
    \ifthenelse{\boolean{figsdir}}{
       \includegraphics*[width=\figsize]{figs/JpsiPiMassFit.eps}
    }{
        \includegraphics*[width=\figsize]{JpsiPiMassFit.eps}
    }
   } 
  \end{center}
  \vskip-0.3cm\caption{\small 
    Invariant-mass distribution of $\bcjp$ candidates (black data points).   
    The maximum likelihood fit of the $B_c^+$ signal is superimposed (blue solid line).
    Individual fit components are also shown: (dashed blue line) the signal, 
    (red long-dashed line) the background and (green dotted line) $B_c^+\to\jpsi K^+$ feeddown.
  \label{fig:mjp}
}
\end{figure}

The signal is represented in the fit by a double-sided Crystal Ball (CB) function\cite{Skwarnicki:1986xj}.
The peak position, the Gaussian mass resolution %($\sigma_{\mjp}=13.5\pm0.7$ \mev) 
and the peak amplitude are 
free parameters in the fit, while the parameters describing small non-Gaussian tails
are fixed by a fit to the simulated signal distribution.  
Using a Gaussian function to model the signal results in a 2.3\%\ relative change in $\R$ value, and this is 
assigned as the systematic uncertainty.
The background is smoothly distributed and modeled by an exponential function.
Varying the background parameterization and the fit range results in up to a 0.6\%\ relative change in $\R$.
A small background from $\bcjk$ decays, peaking $37 \mev$ below the signal peak, 
is also included in the fit with all shape parameters fixed from the simulation. 
Its normalization is constrained to be 1\%\ of the fitted signal amplitude, 
as predicted by the measured ratio of the branching fractions \cite{LHCb-PAPER-2013-021} 
scaled by an efficiency ratio of 15\%\ obtained from the simulation.
The relative systematic uncertainty on $\R$ related to this fit component is 0.1\%.

\section{Extraction of the $\bcjm$ signal} 
\label{sec:bcjm}

To measure the $\Bc\to\jpsi\mu^+\nu_{\mu}$ rate, 
feeddown from other $\Bc\to\Yfeed$, $\Yfeed\to\jpsi\mu^+\nu_{\mu}\,X$ decays 
must be accounted for.
Decays to excited charmonium states 
($\Yfeed=\psifeed\mu^+\nu_{\mu}$, with $\psifeed=\chi_{cJ}$ or $\psi(2S)$)
and states containing $\tau$ leptons ($\Yfeed=\jpsi\tau^+\nu_\tau$) are the dominant contributions.    
Since the rates for such decays have not been measured, 
we rely on theoretical predictions for 
\begin{equation}
\Rf \equiv \frac{\BR(\Bc\to\Yfeed)}{\BR(\Bc\to\jpsi\mu^+\nu_{\mu})}.
\end{equation}
Although the spread in $\Rf$ predictions is large, 
 the related systematic uncertainty is minimized by
restricting the analysis to the high $\jpsi\mu^+$ mass region.
Unreconstructed decay products in 
the $\psifeed\to\jpsi X$ transitions 
($X=\gamma$, $\pi\pi$, $\pi^0$, $\eta$, $\gamma\gamma$) 
or $\tau^+\to\mu^+\nu_\mu \bar{\nu}_\tau$ decays 
carry energy away, lowering the $\jpsi\mu^+$ mass relative to that from direct 
$\Bc\to\jpsi\mu^+\nu_{\mu}$ decays, as illustrated in Fig.~\ref{fig:mjmmc}.
The selection requirement in $\mjm$ is chosen to eliminate the backgrounds 
from $B_{u,d,s}$ decays to $\jpsi$ mesons associated with hadrons, with one of the hadrons misidentified as a muon.
These backgrounds are large because the $B_{u,d,s}$ production rates are orders of magnitude higher than for $B_c^+$. 
Since many exclusive decay modes with various hadron multiplicities and unknown branching ratios contribute, 
the $\mjm$ shape of such backgrounds 
is difficult to predict.
The $5.3 \gev$ lower limit on $\mjm$ 
is above the kinematic limit for $B_u^+\to\jpsi h^+$ decays, with $h^+$ denoting a charged kaon or pion, 
as illustrated in Fig.~\ref{fig:mjmmc}.
\begin{figure}[t]
  \begin{center}
  \ifthenelse{\boolean{pdflatex}}{
    \ifthenelse{\boolean{figsdir}}{
    \includegraphics*[width=\figsize]{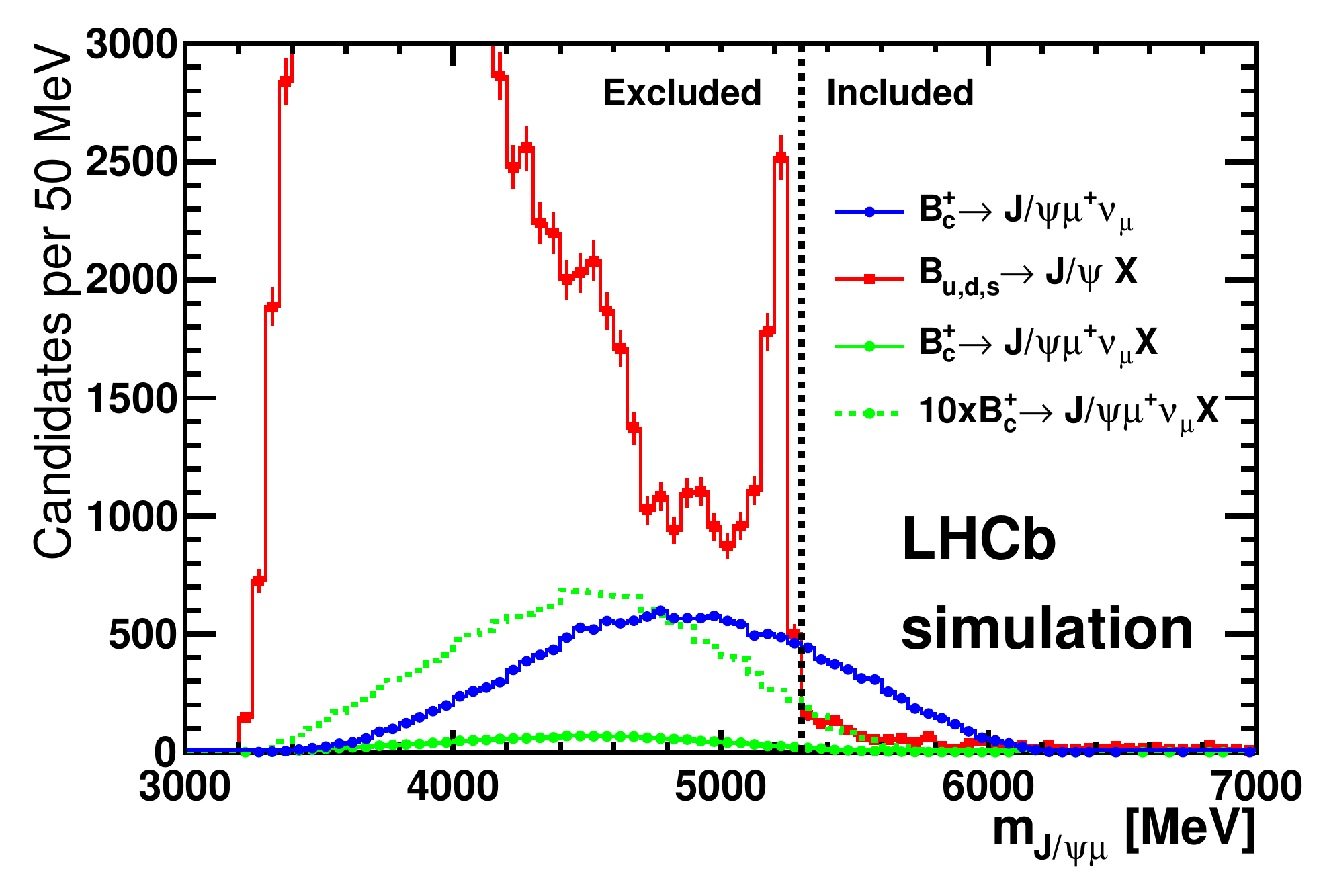}
    }{
    \includegraphics*[width=\figsize]{JpsiMu_MC.pdf}
    }
   }{
    \ifthenelse{\boolean{figsdir}}{
    \includegraphics*[width=\figsize]{figs/JpsiMu_MC.eps}
    }{
    \includegraphics*[width=\figsize]{JpsiMu_MC.eps}
    }
   } 
  \end{center}
  \vskip-0.3cm\caption{\small 
    Distribution of $\mjm$ for $B_c^+\to\jpsi\mu^+\nu_\mu$ candidates 
    selected in simulated event samples of (blue filled points) 
    the signal,
    (green filled points) the $B_c^+$ feeddown
    and (red filled squares) the $B_{u,d,s}$ backgrounds.
    Relative normalization is derived from the fit to the data
    described later in the text.
    The part of the spectrum included in the fit is indicated
    with a vertical dashed black line. 
    The $B_c^+$ feeddown distribution is also shown after
    magnifying its normalization by a factor of ten
    (green dashed histogram).  
  \label{fig:mjmmc}
}
\end{figure}
The $B_{u,d,s}$ backgrounds in the selected region are much smaller, 
and are from $B_{u,d,s}\to\jpsi X$ decays paired with
a bachelor $\mu^+$ originating from a semileptonic decay of the companion $b$ quark in the 
produced $b\bar b$ pair.  
Simulation of $b$-baryon decays to final states involving a $\jpsi$ meson 
shows that they also contribute via this mechanism. 
The shape of such combinatorial backgrounds is less sensitive to the details of the 
composition of $b$-hadron decay modes, and thus is easier to predict. 
Since the combinatorial backgrounds are dominated by genuine muons, the analysis is not 
sensitive to the estimation of muon misidentification rates and associated systematic uncertainties.

The $\mjm$ signal shape is dominated by the endpoint kinematics, whereas
the combinatorial background is smooth and extends beyond the kinematic limit for the $B_c^+\to\jpsi\mu^+\nu_\mu$ decays.
The signal yield is determined by a fit to the $\mjm$ distribution.   
The feeddown background is small as discussed in detail below. 
Its shape is constrained by simulation, while its normalization is related to the 
signal yield via theoretical predictions. 
The unbinned maximum likelihood fit is performed simultaneously to the $\mjm$ distribution
in data and the signal and background distributions from simulation, 
in the range of $5.3$ to $8.0 \gev$, and gives
$N_{\jm}=3537\pm125$ signal events. 
The $\mjm$ distributions and the fit results are displayed in Fig.~\ref{fig:mjmfit}.
The fit is described in detail below. 

\ifthenelse{\boolean{prl}}{
\begin{figure*}[tbhp]
}{
\begin{figure}[t]
}
\begin{center}
  \ifthenelse{\boolean{pdflatex}}{
    \quad
    \hbox{
\ifthenelse{\boolean{figsdir}}{
\ifthenelse{\boolean{prl}}{
    \includegraphics*[width=\figsize]{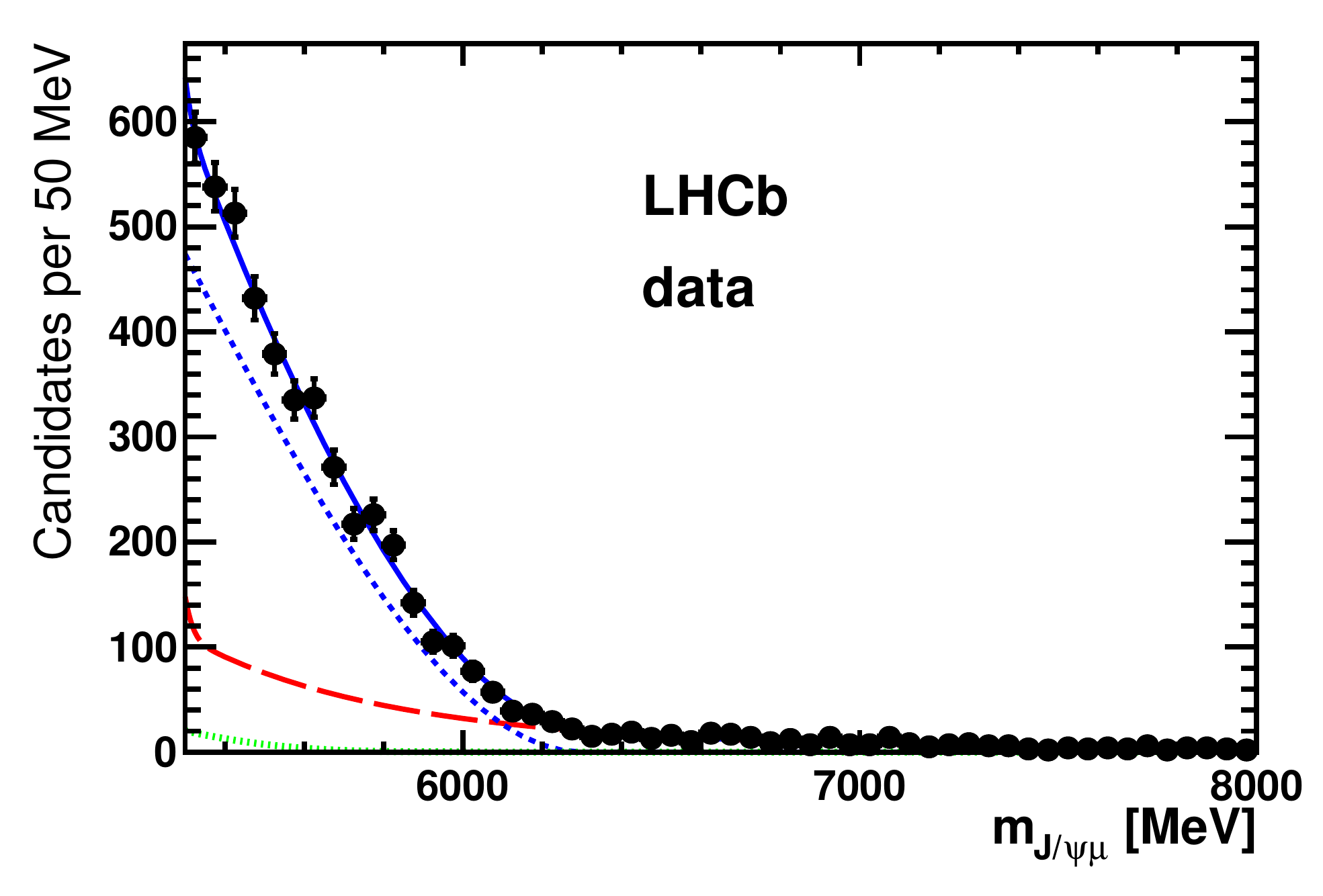} 
    \includegraphics*[width=\figsize]{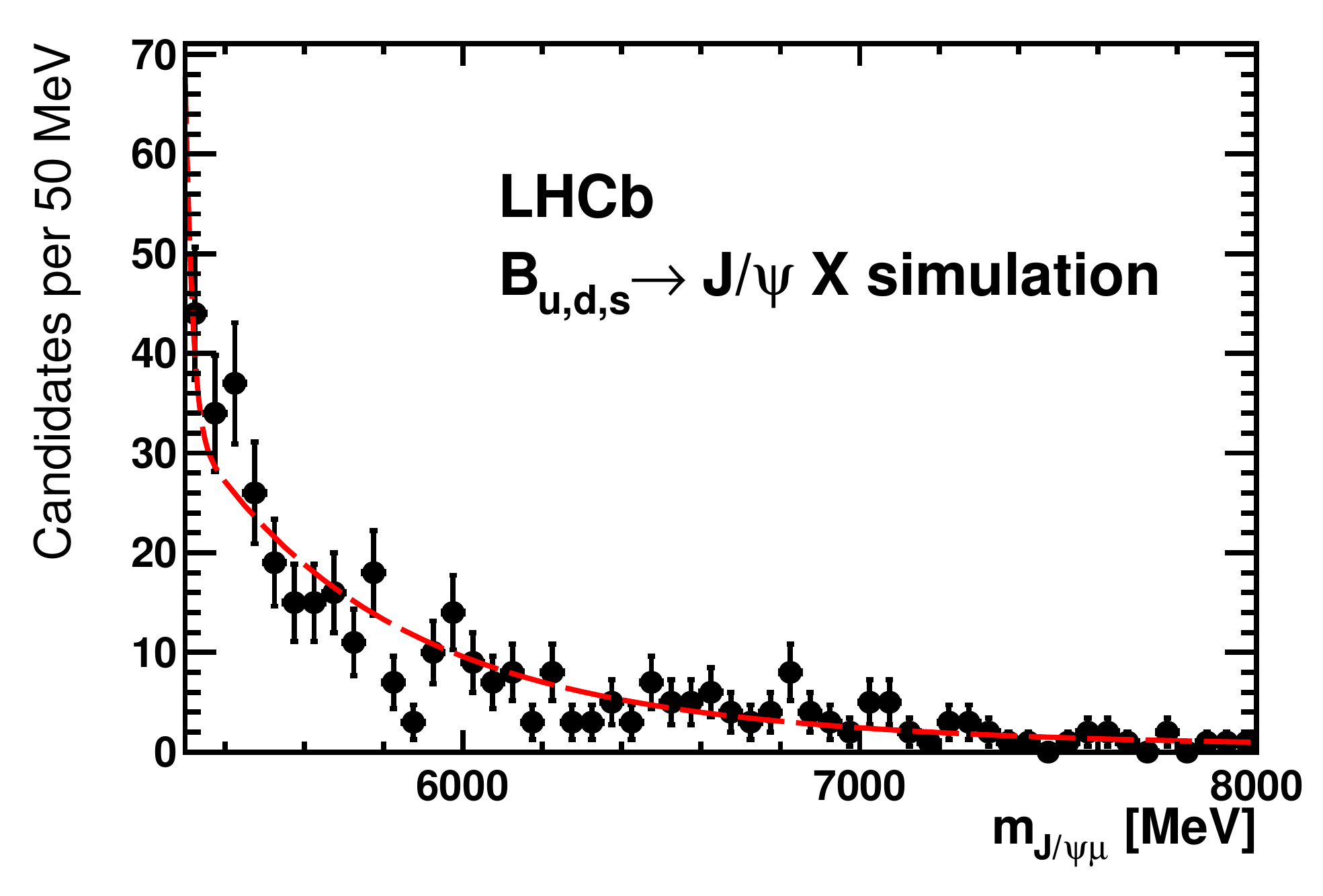} 
}{
    \includegraphics*[width=0.6\figsize]{figs/JpsiMu_Data.pdf} 
    \includegraphics*[width=0.6\figsize]{figs/JpsiMu_BkgMC.pdf} 
}
}{
\ifthenelse{\boolean{prl}}{
    \includegraphics*[width=\figsize]{JpsiMu_Data.pdf} 
    \includegraphics*[width=\figsize]{JpsiMu_BkgMC.pdf} 
}{
    \includegraphics*[width=0.6\figsize]{JpsiMu_Data.pdf} 
    \includegraphics*[width=0.6\figsize]{JpsiMu_BkgMC.pdf} 
}
}
    } \\[-0.5cm]
    \quad
    \hbox{
\ifthenelse{\boolean{figsdir}}{
\ifthenelse{\boolean{prl}}{
    \includegraphics*[width=\figsize]{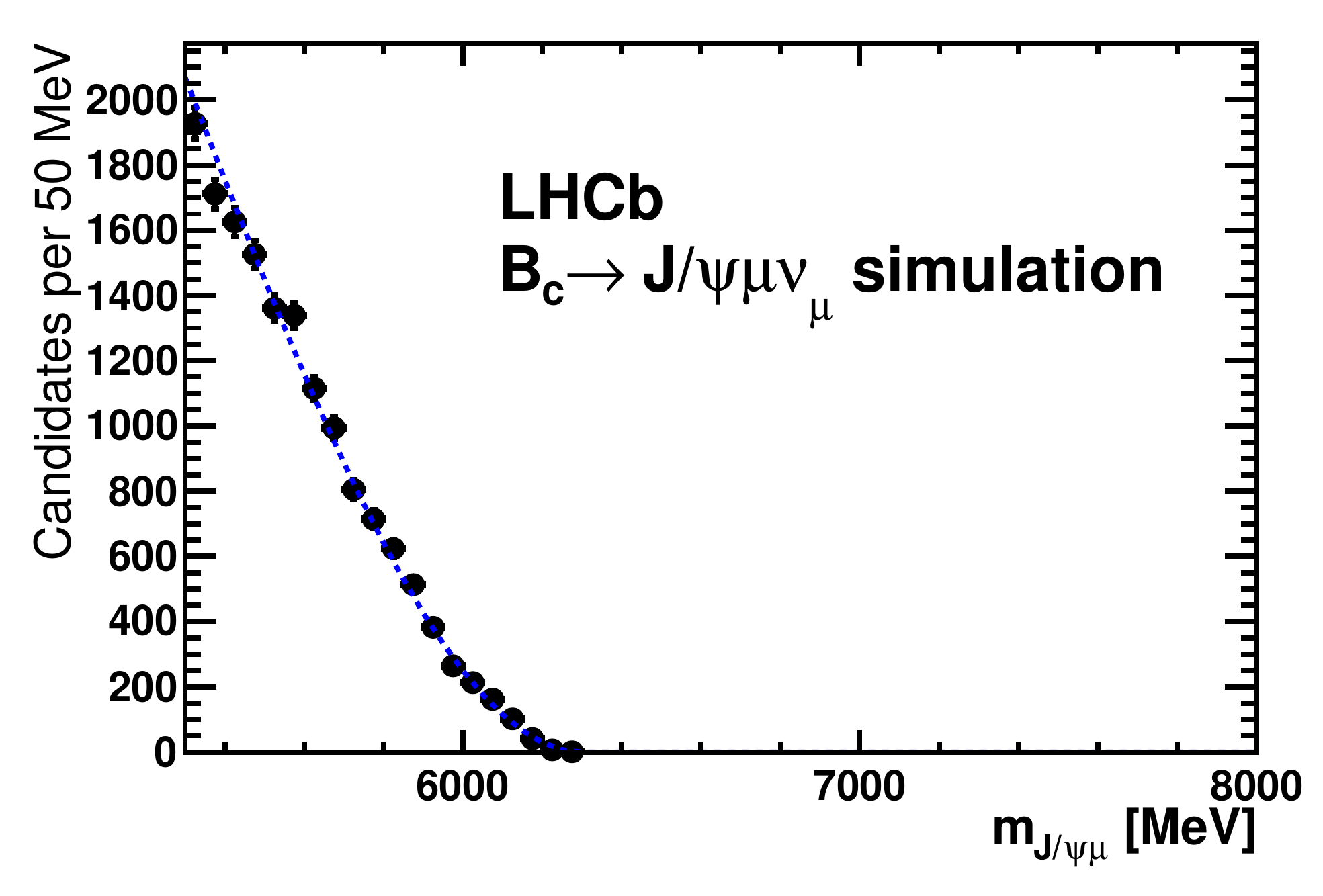}
    \includegraphics*[width=\figsize]{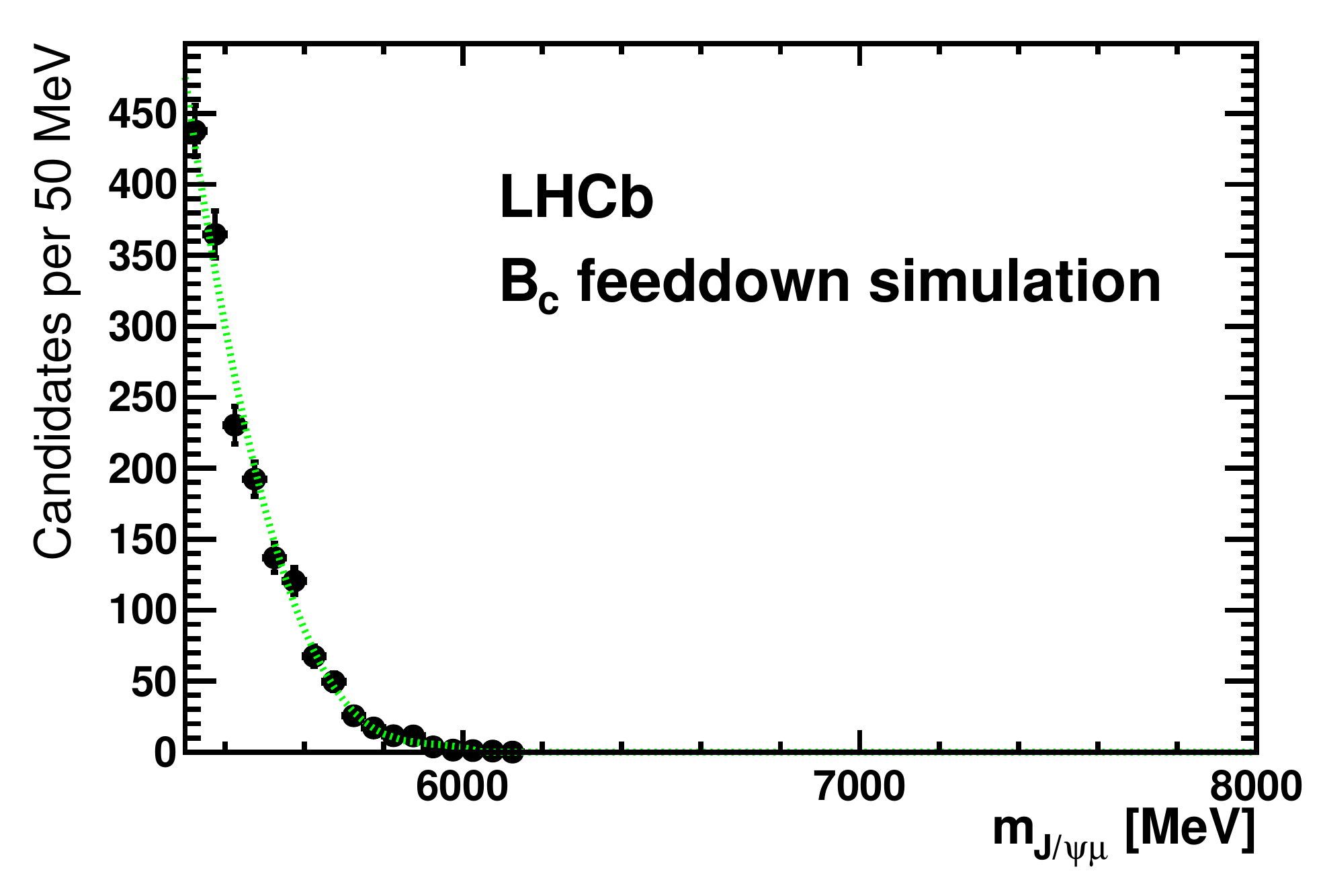} 
}{
    \includegraphics*[width=0.6\figsize]{figs/JpsiMu_SignalMC.pdf}
    \includegraphics*[width=0.6\figsize]{figs/JpsiMu_FeedDown.pdf} 
}
}{
\ifthenelse{\boolean{prl}}{
    \includegraphics*[width=\figsize]{JpsiMu_SignalMC.pdf}
    \includegraphics*[width=\figsize]{JpsiMu_FeedDown.pdf} 
}{
    \includegraphics*[width=0.6\figsize]{JpsiMu_SignalMC.pdf}
    \includegraphics*[width=0.6\figsize]{JpsiMu_FeedDown.pdf} 
}
}
    }
   }{
    \quad
    \hbox{
\ifthenelse{\boolean{figsdir}}{
\ifthenelse{\boolean{prl}}{
    \includegraphics*[width=\figsize]{figs/JpsiMu_Data.eps} 
    \includegraphics*[width=\figsize]{figs/JpsiMu_BkgMC.eps} 
}{
    \includegraphics*[width=0.6\figsize]{figs/JpsiMu_Data.eps} 
    \includegraphics*[width=0.6\figsize]{figs/JpsiMu_BkgMC.eps} 
}
}{
\ifthenelse{\boolean{prl}}{
    \includegraphics*[width=\figsize]{JpsiMu_Data.eps} 
    \includegraphics*[width=\figsize]{JpsiMu_BkgMC.eps} 
}{
    \includegraphics*[width=0.6\figsize]{JpsiMu_Data.eps} 
    \includegraphics*[width=0.6\figsize]{JpsiMu_BkgMC.eps} 
}
}
    } \\[-0.5cm]
    \quad
    \hbox{
\ifthenelse{\boolean{figsdir}}{
\ifthenelse{\boolean{prl}}{
    \includegraphics*[width=\figsize]{figs/JpsiMu_SignalMC.eps}
    \includegraphics*[width=\figsize]{figs/JpsiMu_FeedDown.eps} 
}{
    \includegraphics*[width=0.6\figsize]{figs/JpsiMu_SignalMC.eps}
    \includegraphics*[width=0.6\figsize]{figs/JpsiMu_FeedDown.eps} 
}
}{
\ifthenelse{\boolean{prl}}{
    \includegraphics*[width=\figsize]{JpsiMu_SignalMC.eps}
    \includegraphics*[width=\figsize]{JpsiMu_FeedDown.eps} 
}{
    \includegraphics*[width=0.6\figsize]{JpsiMu_SignalMC.eps}
    \includegraphics*[width=0.6\figsize]{JpsiMu_FeedDown.eps} 
}
}
    }
   } 
\end{center}
  \vskip-0.3cm\caption{\small
    Invariant-mass distribution of $\jpsi\mu^+$ pairs from $\bcjm$ candidates (black data points)
    for (top left) the data, (bottom left) $B_c^+\to\jpsi\mu^+\nu_\mu$ signal simulation, 
    (top right) $B_{u,d,s}\to\jpsi X$ background simulation and
    (bottom right) $B_c^+$ feeddown simulation.
    The unbinned maximum likelihood fit of the $B_c^+$ signal is superimposed (blue solid line).
    Individual fit components are also shown: (blue short-dashed line) the signal, 
    (red long-dashed line) the background and (green dotted line) $B_c^+$ feeddown.
%    A $p$-value of the $\chi^2$ calculated between the data and the fit is $65\%$.
}
\label{fig:mjmfit}
\ifthenelse{\boolean{prl}}{
\end{figure*}
}{
\end{figure}
}

The total PDF used in the fit is the sum of the signal PDF ($\PDF_{\rm sig}$), 
the feeddown background PDF ($\PDF_{\rm fd}$) and the combinatorial background PDF ($\PDF_{\rm bkg}$),
\begin{equation}
\ifthenelse{\boolean{prl}}{\begin{aligned}}{}
\PDF(\mjm) \propto
\ifthenelse{\boolean{prl}}{&}{}
N_{\jm}\,\left( \PDF_{\rm sig}(\mjm) + \alpha\,\PDF_{\rm fd}(\mjm) \right)
\ifthenelse{\boolean{prl}}{\\}{}
+ 
\ifthenelse{\boolean{prl}}{&}{}
N_{\rm bkg}\,\PDF_{\rm bkg}(\mjm), 
\ifthenelse{\boolean{prl}}{\end{aligned}}{}
\end{equation}
where 
$\alpha$ is the feeddown-to-signal yield ratio 
and
$N_{\rm bkg}$ is the combinatorial background yield.
The signal shape is dominated by the endpoint kinematics, thus it is 
modeled as
\begin{equation}
\PDF_{\rm sig}(\mjm) \propto \mathrm{PS}(\mjm) \, (1+s_1\, \tmjm ),
\end{equation}
where $\tmjm=\mjm-5.3 \gev$ and
$\mathrm{PS}(\mjm)$ corresponds to the uniform distribution in 
the $B_c^+\to\jpsi\mu^+\nu_{\mu}$ three-body phase-space,
\begin{equation}
\label{eq:PS}
\ifthenelse{\boolean{prl}}{\begin{aligned}}{}
\mathrm{PS}(\mjm) = 
\ifthenelse{\boolean{prl}}{&}{} 
\frac{{M_{B_c}}^2 - {\mjm}^2}{\mjm} 
\ifthenelse{\boolean{prl}}{\\ \times &}{} 
        \sqrt{ {\mjm}^2-(M_{\jpsi}+M_{\mu})^2 } 
\ifthenelse{\boolean{prl}}{\\ \times &}{} 
        \sqrt{ {\mjm}^2-(M_{\jpsi}-M_{\mu})^2 } 
\ifthenelse{\boolean{prl}}{\end{aligned}}{}
\end{equation}
with the $\jpsi$ and $\mu$ masses ($M_{\jpsi}$ and $M_\mu$) set to their known values \cite{PDG},
and $M_{B_c}$ set to an effective value, % $6.289$ \gev, 
which is slightly higher than the $B_c^+$ mass to account for  
detector resolution effects.
Setting $M_{B_c}$ to the known $B_c^+$ mass \cite{PDG} changes the signal yield by a negligible amount.
Deviations from the uniform distribution are allowed by the linear term, with the $s_1$ 
coefficient determined by the simultaneous fit to the simulated signal distribution and the data.
The simulation based on the Kiselev \etal~QCD sum rules model \cite{Kiselev} is used in the default fit.
The models of Ebert \etal \cite{Ebert}, based on a relativistic quasipotential Schr\"odinger approach,
and ISGW2 \cite{ISGW2}, 
based on a nonrelativistic constituent quark model with relativistic corrections,
alter the determined signal yield by $+0.2\%$ and
$-0.4\%$, respectively.
Relying on the data themselves to determine the signal shape
changes the signal yield by $+0.7\%$. The latter value is taken as a systematic error.         

The feeddown includes contributions from the following $B_c^+$ decay modes 
$f=\psi(2S)\mu^+\nu_\mu$, 
$\chi_{cJ}\mu^+\nu_\mu$
and
$\jpsi\tau^+\nu_\tau$.
Feeddown from $B_c^+\to B_{d,s} \mu^+\nu_\mu$ and $B_c^+\to\jpsi$ plus hadrons
is also investigated and found negligible.
Their individual proportions with respect to the signal yield are determined as
\begin{equation}
\alpha_f = \Rf \, \Bcasc \, \Reps,
\end{equation}
and then added, $\alpha=\sum_f \alpha_f $,
where $\Bcasc$ is
the sum of the measured branching fractions \cite{PDG} for the $\psifeed$ state to 
decay to a $\jpsi$ meson by emission of unreconstructed photons or light hadrons,
%\begin{equation}
%\Bcasc=\sum_X \BR(\psifeed\to \jpsi X),
%\end{equation}
and $\Reps$ is the ratio of the feeddown and the signal reconstruction 
efficiencies.\footnote {For $B_c^+\to\jpsi\tau^+\nu_\tau$, 
$\Bcasc=\BR(\tau^+\to\mu^+\nu_{\mu}\bar{\nu}_\tau)$.}
This quantity is small because of the $\mjm>5.3 \gev$ requirement.
For $\chi_{cJ}$ states the sum extends over the three $J$ values,
$\Rf \, \Bcasc = \sum_{J=0,1,2} \RfJ \, \BcascJ$.   
The values of the parameters affecting the estimate of the feeddown fraction are summarized
in Table~\ref{tab:feeddown}.
Theoretical predictions
for $\Rf$ for the $B_c^+\to\psi(2S)\mu^+\nu_\mu$ feeddown mode 
vary over a wide range,
%(0.009 \cite{RefLC},
% 0.025 \cite{Ebert},
% 0.042 \cite{Ref3CC},
% 0.050 \cite{Kiselev}, 
% 0.080 \cite{Ref5CF},
% 0.090, 0.164 \cite{RefKLL},
% 0.185 \cite{ISGW2}),
0.009--0.185 \cite{RefLC,Ebert,Ref3CC,Kiselev,Ref5CF,RefKLL,ISGW2}.
An average of the highest and the lowest prediction is taken for the nominal estimate,
and half of the difference is taken for the systematic error.  
The theoretical uncertainties in the $\Rf \, \Bcasc$ values 
for the dominant $B_c^+\to\chi_{cJ}\mu^+\nu_\mu$ feeddown mode are smaller,
%($0.032$ \cite{Ref1IKS}, $0.038$ \cite{Ref2CCWZ,*Ref2CC} and $0.032$ \cite{Ref3HNV}).
0.032--0.038 \cite{Ref1IKS,Ref2CCWZ,Ref3HNV}.
The spread is also limited for theoretical predictions of 
$\Rf$ for the $B_c^+\to\jpsi\tau^+\nu_\tau$ decay,
%($0.237$ \cite{Ref1IKS}, $0.253$ \cite{Kiselev},  $0.266$ \cite{Ref3HNV} and $0.283$ \cite{Ref4ANSS1,*Ref4AKNT2}).
0.237--0.283 \cite{Ref1IKS,Kiselev,Ref3HNV,Ref4ANSS1,*Ref4AKNT2}.
The simulated distributions for the
individual feeddown modes are mixed
according to the proportions resulting from the $\Rf\,\Bcasc$ values
and then parameterized as
\begin{equation}
\PDF_{\rm fd}(\mjm) \propto \mathrm{PS}(\mjm) \, (1+f_1\, \tmjm+f_2\, {\tmjm}^2),
\end{equation}
where $f_1$ and $f_2$ are parameters determined by the fit. 
The effect of the unreconstructed decay products $X$ is to lower the 
effective $M_{B_c}$ value in Eq.~(\ref{eq:PS}).
Varying the feeddown fraction within its uncertainty changes the signal yield by up to $0.6\%$.

\ifthenelse{\boolean{prl}}{
\begin{table*}[tbh]
}{
\begin{table}[t]
}
  \caption{\small Values of the parameters affecting the estimate of the feeddown fraction in the fit
         to the $\mjm$ distribution. 
         For $B_c^+\to\chi_{cJ}\mu^+\nu_\mu$, $\sum_{J=0,1,2} \RfJ \, \BcascJ$ is listed. 
}
\label{tab:feeddown}
\centering
\begin{tabular}{lcccc}
\hline
Feeddown mode       &  $\Rf$   & $\Bcasc$  &  $\Reps$ & $\alpha_f$ \\
\hline
$B_c^+\to\psi(2S)\mu^+\nu_\mu$
                    &  $0.009-0.185$       & $0.598\pm0.006$   & $0.118\pm0.004$ & $0.0069\pm0.0062$ \\
$B_c^+\to\chi_{cJ}\mu^+\nu_\mu$ 
                    & \multicolumn{2}{c}{$0.032-0.038$}         & $0.364\pm0.009$ & $0.0127\pm0.0011$ \\
$B_c^+\to\jpsi\tau^+\nu_\tau$ 
                    & $0.237-0.283$         & $0.1741\pm0.0004$ & $0.014\pm0.001$ & $0.0006\pm0.0001$ \\
\hline
Total $\alpha $              &                       &                   &                 & $0.0202\pm0.0063$ \\
\end{tabular}
\ifthenelse{\boolean{prl}}{
\end{table*}
}{
\end{table}
}

The combinatorial $B_{u,d,s}$ background is parameterized with an exponential function.
The tail of the $B_u^+\to\jpsi h^+$ distribution, with the light hadron misidentified as a muon, may enter the
signal region because of detector resolution. 
We parameterize it with a Gaussian function, $G(\mjm)$, with a mean value and 
width fixed to the results of the fit to the simulated $B_u^+\to\jpsi h^+$ distribution. 
The exponential and $G(\mjm)$ functions together define $\PDF_{\rm bkg}(\mjm)$,
\begin{equation*}
\PDF_{\rm bkg}(\mjm) \propto c\,N_e e^{b_1 \tmjm + b_2 {\tmjm}^2} + (1-c)\,G(\mjm),
\end{equation*} 
where $N_e$ normalizes the exponential function to one. 
The combinatorial background fraction $c$ and the polynomial coefficients $b_1$ and $b_2$ are 
free parameters in the simultaneous fit to the simulated 
$B_{u,d,s}\to\jpsi X$ distribution and to the distribution in the data.
To avoid relying on simulation for the absolute values of the muon misidentification rates, 
$c$ is allowed to vary independently in the fit to the simulated and the observed distributions.
A systematic uncertainty of 1.8\%\ is assigned to this background parameterization 
based on fit results in which either the Gaussian term is neglected or 
the exponential function is replaced by a sum of two exponential functions.
 
Varying the upper limit of the mass range used in the fit from 
$8.0$ down to $6.75 \gev$, results in a signal yield change
of up to $1.5\%$. 
Varying the corresponding lower limit from its default value of $5.3$ to $5.1 \gev$, thus 
including the peak of the $B_u^+\to\jpsi h^+$ component (see Fig.~\ref{fig:mjmmc}),
or to $5.5 \gev$, thus avoiding the tail of that component,
results in a relative change in the $\R$ value of up to $1.6\%$.

The default method of the $\bcjm$ signal-yield determination relies on 
simulation to predict the signal and background shapes in the $\mjm$ distribution.
An alternative approach relies on simulation to predict the signal and 
background shapes of the $\DLL$ distribution. 
Correlations between $\mjm$ and $\DLL$ variables are small.
The requirement on the $\DLL$ value is removed. The $\mjm$ range is restricted to 5.3--6.1 $\gev$ 
to exclude the backgrounds above the $B_c^+$ kinematic limit.
The signal and combinatorial background yields are determined by a fit to the $\DLL$ 
distribution in the data. 
The $B_c^+$ feeddown simulation predicts a similar $\DLL$ shape as for the 
$\bcjm$ signal. Therefore, this contribution is not represented explicitly in the fit to
the $\DLL$ distribution, but is subtracted from the fitted signal yield according to the 
feeddown fraction $\alpha$. 
Taking into account the differences in signal efficiency, the $\bcjm$ signal yield is
consistent with that resulting from the $\mjm$ fit method within $0.5\%$, which is 
included as an additional systematic uncertainty 
due to the $\DLL$ requirement in the nominal approach.

\section{Results}

The ratio of the reconstruction efficiencies 
between the two $B_c^+$ signal modes, as determined from 
simulation, is $\epsilon(\bcjm)/\epsilon(\bcjp)=1.14\pm0.01$ (statistical error) for 
$\bcjm$ events generated in the endpoint region. 
Using different $\bcjm$ form factor models changes this efficiency 
ratio by up to $1.3\%$.
Efficiencies of the 
pion and muon particle identification (PID) 
requirements have systematic uncertainties of $0.8\%$ and $1.9\%$, respectively.
The efficiency-ratio systematic uncertainties from the $B_c^+$ lifetime assumed in
the simulation is $0.2\%$ due to the cancelations between the two 
decay modes. 
The fraction of multiple signal candidates per event is $0.1\%$ for $\bcjp$ and
$1.9\%$ for $\bcjm$ decays. 
To check for possible biases due to the neglected correlations between multiple candidates, 
one candidate is randomly chosen, which changes the $\R$ result by $0.4\%$.
The systematic uncertainty associated with the limited knowledge of 
the efficiency of the $\DLL$ requirement for              
$\bcjm$ decays is included using the results of the $\DLL$ fit. 
To study the corresponding uncertainty for $\bcjp$ decays, the $\DLL$ requirement is varied, resulting 
in a $2\%$ variation. 
The systematic uncertainty associated with the trigger simulation is $3.4\%$, 
as estimated by modifying the trigger requirements.
The systematic errors are summarized in Table~\ref{tab:sys}.
The total relative systematic uncertainty on $\R(\mjm>5.3 \gev)$ is $6\%$.

\ifthenelse{\boolean{prl}}{
\begin{table}[hbt]
}{
\begin{table}[t]
}
 \caption{\small Summary of systematic uncertainties. The total systematic errors are obtained by adding in quadrature
                 the individual contributions.}
\begin{center}
\begin{tabular}{lr}
    Contribution                               & Relative \\
                                               & error    \\
\hline
    $\mjp$ signal shape                 & $2.3\%$   \\
    $\mjp$ background shape             & $0.2\%$  \\       
    $B_c^+\rightarrow\jpsi K^+$ component   & $0.1\%$ \\           
\hline
    $\mjm$ signal shape                 & $0.7\%$   \\
    $\mjm$ background shape             & $1.8\%$     \\ 
    $B_c^+$ feeddown                    & $0.6\%$  \\
    Lower $\mjm$ fit range limit        & $1.6\%$   \\
    Upper $\mjm$ fit range limit        & $1.5\%$   \\    
\hline
    $\bcjm$ model dependence of efficiency     & $1.3\%$ \\        
    Pion PID                                   & $0.8\%$   \\ 
    Muon PID                                   & $1.9\%$   \\ 
    Lifetime                                   & $0.2\%$   \\     
    Multiple candidates                        & $0.4\%$  \\         
    $\DLL$ requirement for $\bcjp$             & $2.0\%$  \\             
    $\DLL$ requirement for $\bcjm$             & $0.5\%$  \\
    Trigger simulation                         & $3.4\%$ \\ 
\hline 
    Total within selected $\mjm$ range     & $6.0\%$     \\
\hline
    $\mjm$ extrapolation                   & $7.9\%$ \\ 
\hline
    Total                                      & $9.9\%$ \\                 
\end{tabular}
\end{center}
\label{tab:sys}
\end{table}

The result for the ratio of the branching fractions 
restricted to decays with $\mjm>5.3 \gev$ is
\begin{equation}
\R(\mjm>5.3 \gev)=0.271\pm0.016\pm0.016,
\end{equation}
where the first uncertainty is statistical and the second is systematic.
This ratio is extrapolated to the full phase-space as follows.
The model of Kiselev \etal\ \cite{Kiselev} predicts the fraction of the
$\bcjm$ rate with $\mjm$ above $5.3 \gev$ to be $0.173$, which is close to
an average over six different models \cite{RefWSL,RefKLL,RefWFX,Ebert,ISGW2}.
The largest deviation from this prediction is $7.9\%$, which is taken
as an estimate of the extrapolation systematic error. 
This increases the systematic uncertainty on $\R$, when extrapolated to the
full mass range, to $9.9\%$ yielding
\begin{equation}
\R = 0.0469 \pm  0.0028 \pm 0.0046.
\end{equation}

A comparison between the measured and the predicted values of $\R$ is 
shown in Fig.~\ref{fig:models}.
The measured value is slightly below the lowest predicted value.
The predictions by the relativistic quasipotential Schr\"odinger model of Ebert \etal \cite{Ebert} 
and the model of El-Hady \etal, based on a nonrelativistic reduction of the Bethe-Salpeter equation \cite{Ref7EMV}, 
are in good agreement with the experimental value. 
The model of Ke \etal \cite{RefKLL}, based on the modified harmonic oscillator wave function in 
light-front quark model, is also consistent with the data.
The other models \cite{Ref3CC,Ref4ANSS1,*Ref4AKNT2,Ref5CF,Kiselev,Ref1IKS}
significantly overestimate $\R$.

\ifthenelse{\boolean{prl}}{
\begin{figure}[tbhp]
}{
\begin{figure}[t]
}
  \begin{center}
  \ifthenelse{\boolean{pdflatex}}{
    \ifthenelse{\boolean{figsdir}}{
    \includegraphics*[width=\figsize]{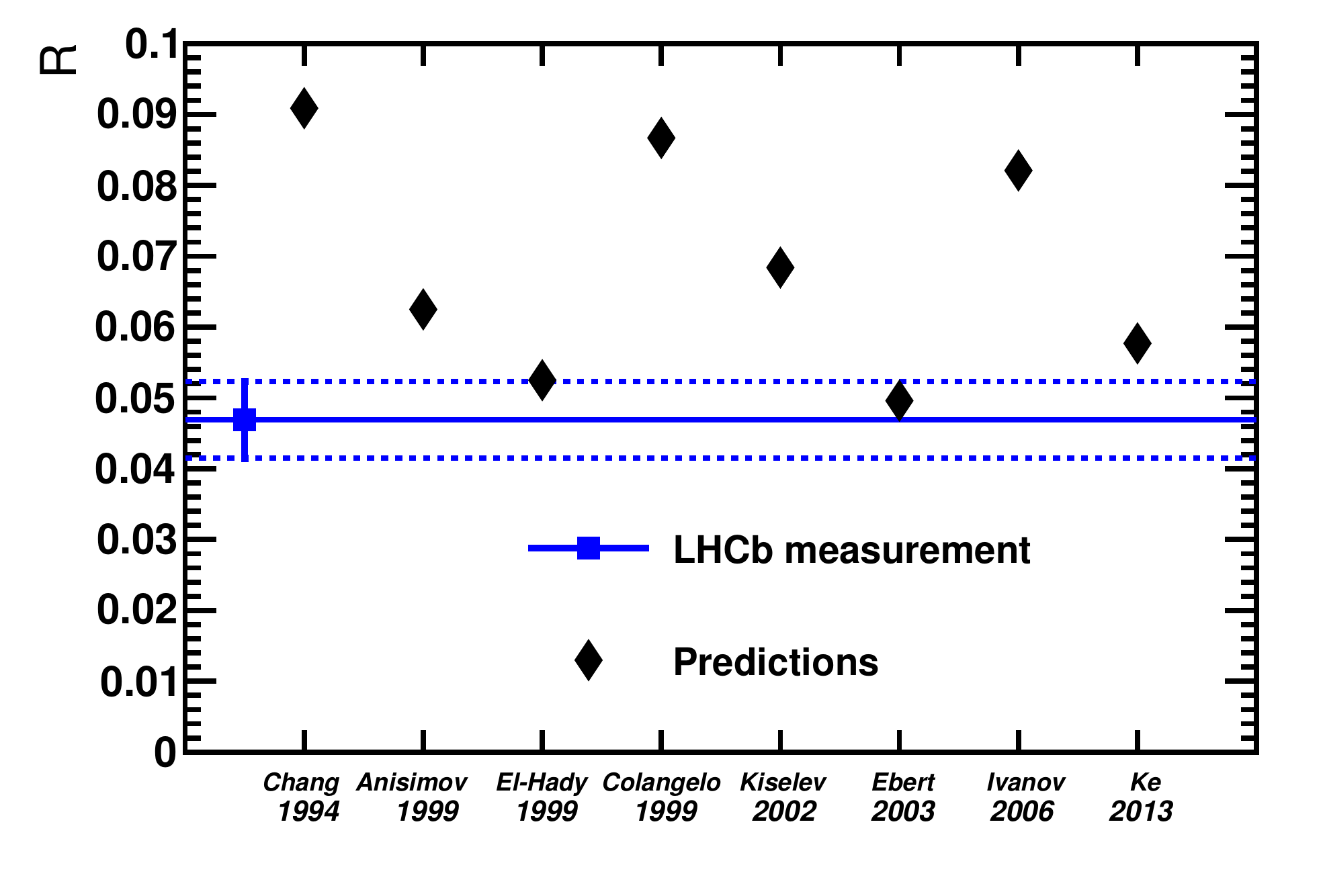}
    }{
    \includegraphics*[width=\figsize]{TheoryCompPlot.pdf}
    }
   }{
    \ifthenelse{\boolean{figsdir}}{
    \includegraphics*[width=\figsize]{figs/TheoryCompPlot.eps}
    }{
    \includegraphics*[width=\figsize]{TheoryCompPlot.eps}
    }
   } 
  \end{center}
  \vskip-0.3cm\caption{\small 
    The measured value of $\R$ (horizontal solid line) and its $\pm1\sigma$ 
    uncertainty band (dashed lines) compared to the predictions (diamonds).
    A nonrelativistic reduction of the Bethe-Salpeter equation is used in the predictions of Chang \etal~\cite{Ref3CC}, 
    El-Hady \etal\cite{Ref7EMV}, and Colangelo \etal\cite{Ref5CF}, while the latter also utilizes
    heavy quark symmetry. A light-front constituent quark model is used by Anisimov \etal\cite{Ref4ANSS1,*Ref4AKNT2}
    and Ke~\etal\cite{RefKLL}. QCD sum rules are used by Kiselev \etal\cite{Kiselev},
    a relativistic quasipotential Schr\"odinger model is used by Ebert \etal\cite{Ebert}, and 
    a relativistic constituent quark model in used by Ivanov \etal\cite{Ref1IKS}.    
  \label{fig:models}
}
\end{figure}

\section{Summary}

The ratio of hadronic and semileptonic decay branching fractions
of the $B_c^+$ meson is measured for the first time. 
Within the observed mass range, $m_{\jpsi\mu}>5.3 \gev$,
the measured value of $\BR(\bcjp)/\BR(\bcjm)$ is found to be $0.271 \pm  0.016\,\stat \pm 0.016 \,\syst$. 
Extrapolating to the full mass range, we obtain a value of
$\BR(\bcjp)/\BR(\bcjm)=0.0469 \pm  0.0028\,\stat \pm 0.0046\,\syst$, which is in
good agreement with 
the theoretical predictions by Ebert \etal \cite{Ebert} and El-Hady \etal \cite{Ref7EMV},
and consistent with the prediction by 
Ke \etal \cite{RefKLL}.
All other currently available models \cite{Ref3CC,Ref4ANSS1,*Ref4AKNT2,Ref5CF,Kiselev,Ref1IKS}
overestimate this ratio.

\section*{Acknowledgements}

\noindent We express our gratitude to our colleagues in the CERN
accelerator departments for the excellent performance of the LHC. We
thank the technical and administrative staff at the LHCb
institutes. We acknowledge support from CERN and from the national
agencies: CAPES, CNPq, FAPERJ and FINEP (Brazil); NSFC (China);
CNRS/IN2P3 (France); BMBF, DFG, HGF and MPG (Germany); SFI (Ireland); INFN (Italy); 
FOM and NWO (The Netherlands); MNiSW and NCN (Poland); MEN/IFA (Romania); 
MinES and FANO (Russia); MinECo (Spain); SNSF and SER (Switzerland); 
NASU (Ukraine); STFC (United Kingdom); NSF (USA).
The Tier1 computing centres are supported by IN2P3 (France), KIT and BMBF 
(Germany), INFN (Italy), NWO and SURF (The Netherlands), PIC (Spain), GridPP 
(United Kingdom).
We are indebted to the communities behind the multiple open 
source software packages on which we depend. We are also thankful for the 
computing resources and the access to software R\&D tools provided by Yandex LLC (Russia).
Individual groups or members have received support from 
EPLANET, Marie Sk\l{}odowska-Curie Actions and ERC (European Union), 
Conseil g\'{e}n\'{e}ral de Haute-Savoie, Labex ENIGMASS and OCEVU, 
R\'{e}gion Auvergne (France), RFBR (Russia), XuntaGal and GENCAT (Spain), Royal Society and Royal
Commission for the Exhibition of 1851 (United Kingdom).

\bibliographystyle{LHCb}
\bibliography{main,LHCb-PAPER,LHCb-CONF,LHCb-DP}

\newpage

% Author List ----------------------------
%  You need to get a new author list!
%%%%%%%%%%%%%%%%%%%%%%%%%%%%%%%%%%%%%%%%%%
\centerline{\large\bf LHCb collaboration}
\begin{flushleft}
\small
R.~Aaij$^{41}$, 
B.~Adeva$^{37}$, 
M.~Adinolfi$^{46}$, 
A.~Affolder$^{52}$, 
Z.~Ajaltouni$^{5}$, 
S.~Akar$^{6}$, 
J.~Albrecht$^{9}$, 
F.~Alessio$^{38}$, 
M.~Alexander$^{51}$, 
S.~Ali$^{41}$
% ...

{\footnotesize \it
% ...
$ ^{5}$Clermont Universit\'{e}, Universit\'{e} Blaise Pascal, CNRS/IN2P3, LPC, Clermont-Ferrand, France\\
$ ^{6}$CPPM, Aix-Marseille Universit\'{e}, CNRS/IN2P3, Marseille, France\\
% ...
$ ^{9}$Fakult\"{a}t Physik, Technische Universit\"{a}t Dortmund, Dortmund, Germany\\
% ...
$ ^{37}$Universidad de Santiago de Compostela, Santiago de Compostela, Spain\\
$ ^{38}$European Organization for Nuclear Research (CERN), Geneva, Switzerland\\
% ...
$ ^{41}$Nikhef National Institute for Subatomic Physics, Amsterdam, The Netherlands\\
% ...
$ ^{46}$H.H. Wills Physics Laboratory, University of Bristol, Bristol, United Kingdom\\
% ...
$ ^{51}$School of Physics and Astronomy, University of Glasgow, Glasgow, United Kingdom\\
$ ^{52}$Oliver Lodge Laboratory, University of Liverpool, Liverpool, United Kingdom\\
% ...
}
\end{flushleft}
%%%%%%%%%%%%%%%%%%%%%%%%%%%%%%%%%%%%%%%%%%

\end{document}